\documentclass{article}

\usepackage{arxiv}

\usepackage[utf8]{inputenc} 
\usepackage[T1]{fontenc}    
\usepackage{amsmath}
\usepackage{hyperref}       
\usepackage{url}            
\usepackage{booktabs}       
\usepackage{amsfonts}       
\usepackage{nicefrac}       
\usepackage{microtype}      
\usepackage{cleveref}       
\usepackage{graphicx}
\usepackage[numbers]{natbib}
\usepackage{doi}
\usepackage{subcaption} 
\usepackage{upgreek} 
\usepackage{color}
\usepackage{tabularx}
\usepackage{makecell} 
\usepackage{cprotect} 
\usepackage{siunitx}

\newcommand{\diff}{\mathrm{d}}

\title{Rarefied gas flow in functionalized microchannels}

\date{January 29, 2023}

\author{Simon Kunze$^1$, Pierre Perrier$^2$, Rodion Groll$^{3,6}$, Benjamin Besser$^1$, \\
\textbf{Stylianos Varoutis$^4$, Andreas Lüttge$^5$, Irina Graur$^2$, Jorg Thöming$^{1,6}$}
\\ \\
\textit{$^1$ University of Bremen, Chemical Process Engineering CVT,} \\ 
\textit{Leobener Str. 6, 28359 Bremen, Germany}\\
\textit{$^2$ Aix Marseille Universit\'e, CNRS, IUSTI UMR 7343, 13453, Marseille, France} \\
\textit{$^3$ University of Bremen, Center of Applied Space Technology and Microgravity}, \\
\textit{Am Fallturm 2, 28359 Bremen} \\
\textit{$^4$ Karlsruhe Institute of Technology KIT, Hermann-von-Helmholtz-Platz 1,} \\
\textit{76344 Eggenstein-Leopoldshafen, Germany} \\
\textit{$^5$ University of Bremen, Center for Marine Environmental Sciences MARUM,} \\ \textit{Klagenfurter Str. 2-4, 28359 Bremen} \\
\textit{$^6$ University of Bremen, MAPEX Center for Materials and Processes,} \\
\textit{Postfach 330 440, 28334 Bremen}
}

\renewcommand{\shorttitle}{Rarefied gas flow in functionalized microchannels}

\begin{document}
\maketitle

\begin{abstract}
	The interaction of rarefied gases with functionalized surfaces is of great importance in technical applications such as gas separation membranes and catalysis. To investigate the influence of functionalization and rarefaction on gas flow rate in a defined geometry, pressure-driven gas flow experiments with helium and carbon dioxide through plain and alkyl-functionalized microchannels are performed. The experiments cover Knudsen numbers from 0.01 to 200 and therefore the slip flow regime up to free molecular flow. To minimize the experimental uncertainty which is prevalent in micro flow experiments, a methodology is developed to make optimal use of the measurement data. The results are compared to an analytical model predicting rarefied gas flow in straight channels and to numerical simulations of the S-model and BGK equations. The experimental data shows no significant difference between plain and functionalized channels. This stands in contrast to previous measurements in smaller geometries and demonstrates that the surface-to-volume ratio seems to be too small for the functionalization to have an influence and highlights the importance of geometric scale for surface effects. These results also shed light on the molecular reflection characteristics described by the TMAC.
\end{abstract}

\section{Introduction}

When the mean free path of gas molecules approaches the smallest size of the surrounding geometry, i.e. pore or channel diameter, the gas is considered to be in a rarefied state. This state is of relevance for many scientific and technical applications, for example gas separation membranes \cite{Baker2012}, catalysis \cite{Montessori2016}, vacuum \cite{Jousten2008} and space technologies \cite{White2013}. Gas behaves differently in a rarefied state and cannot be described by the continuum approach, therefore more general kinetic methods need to be used \cite{Shen2005}. In addition, under rarefied conditions the interaction between gas molecules and surface is enhanced which can strongly influences the gas flow. This effect is utilized to increase selectivity in membrane applications. A surface can be functionalized by attaching molecules with specific functional groups onto the surface which selectively interact with the gas molecules.

However, previous experimental studies have shown that applying such surface functionalization significantly reduces the gas flow \cite{Leger1996,Stoltenberg2012,Suzuki2014}. It was shown that length and density of the functional molecules are the determining factors which impact the flow reduction \cite{Besser2017,Besser2020}. The chemical composition of the functional group itself does not influence the gas flow.

To investigate this behavior in a larger and, compared to porous media, well-defined geometry, pressure-driven gas flow experiments in straight rectangular microchannels are performed. Hexadecyltrimethoxysilane (HDTMS), a silane molecule with a C$_{16}$ chain as functional group, is deposited onto the channel surfaces by chemical vapor deposition and its influence on the gas flow is investigated.

The experimental results are analyzed for differences in mass flow rate between plain and functionalized channels and the data is compared to numerical solutions of the linearized S-model kinetic equation \cite{Graur2014}, the linearized Bhatnagar-Gross-Krook (BGK) kinetic equation \cite{Varoutis2009}, and to an analytical model proposed in \cite{Kunze2022}.

\section{Methods}

The mass flow measurements are performed with the microchannels described below in a bench-scale setup at the IUSTI laboratory in Marseille, France and in the TRANSFLOW facility at the KIT laboratory in Karlsruhe, Germany. The microchannels are prepared at CVT and MARUM in Bremen, Germany.

\subsection{Microchannel characteristics}
\label{subsec:micro_channels}
\subsubsection{Geometry and socket}
\label{subsubsec:micro_channel_geometry_socket}
The microchannels are manufactured by wet etching on silicon wafers, with the channels aligned parallel to the wafer surface. The etching depth corresponds to the channel height and the width is fixed by the masking of the wafer. After anodic bonding with borosilicate glass to close the channels, the length is set by cutting the wafer at specific points.

Before the bonding process, the width and height of the channels are measured via vertical scanning interferometry (VSI). VSI is an optical, non-destructive method (e.g., \cite{Luttge1999,Arvidson2014}) that uses white light in the present application. The vertical resolution of the instrument is typically about 1 nanometer, the lateral resolution varies depending on the (mirau) objectives used. The length of the channels is measured using a caliper gauge. Two different channel geometries are used. For the mass flow measurements at the IUSTI laboratory, a stack of 100 parallel channels is manufactured, of which 99 channels are closed by using epoxy glue (\textit{UHU Plus Endfest}) to perform single-channel measurements. For the KIT laboratory, 20 channels in parallel are used for increasing the mass flow to meet the facility's requirements. The geometries of the channels are given in Table \ref{tab:channels}.

\begin{table}
    \caption{Channel characteristics.}
    \centering
    \begin{tabular}{@{}llllll}
    \toprule
    Channel &  & Height h & Width w  & Length L & Parallel \\
    size & Laboratory & [$\upmu$m] & [$\upmu$m] & [mm] & channels\\
    \midrule
    small & IUSTI & 5.21 $\pm$ 0.1  & 145.22 $\pm$ 0.21 & 12.07 $\pm$ 0.06 & 1 \\
    large & KIT & 48.2 $\pm$ 0.3 & 1469 $\pm$ 6 & 12.56 $\pm$ 0.06 & 20 \\
    \bottomrule
    \end{tabular}
    \label{tab:channels}
\end{table}

\begin{figure}[ht]
    \centering
    \begin{subfigure}{.45\textwidth}
        \centering
        \includegraphics[width=\linewidth]{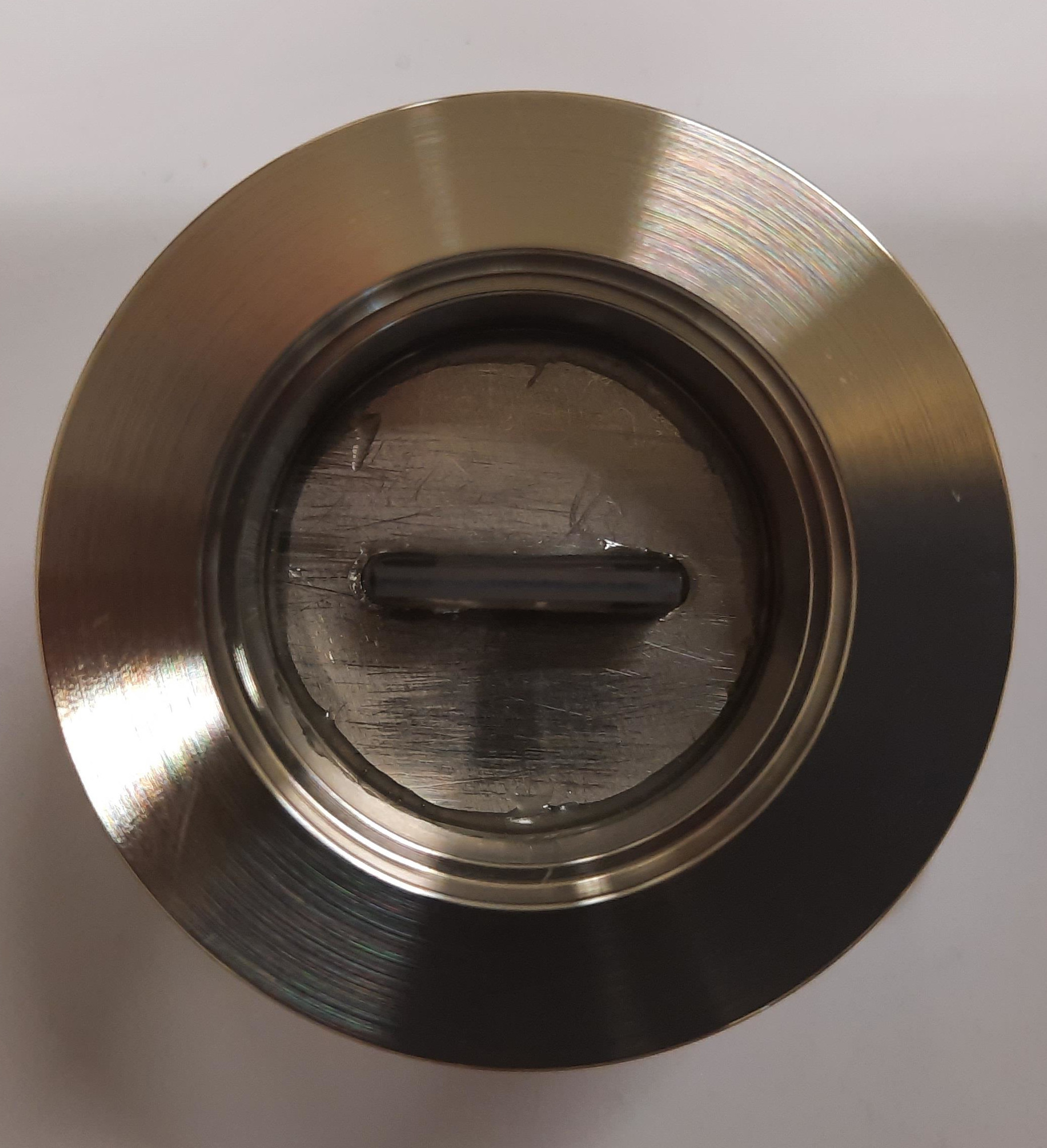}
        \caption{ }
        \label{fig:KF_flange} 
    \end{subfigure}
    \hspace*{0.02\textwidth}
    \begin{subfigure}{.45\textwidth}
        \centering
        \includegraphics[width=\linewidth]{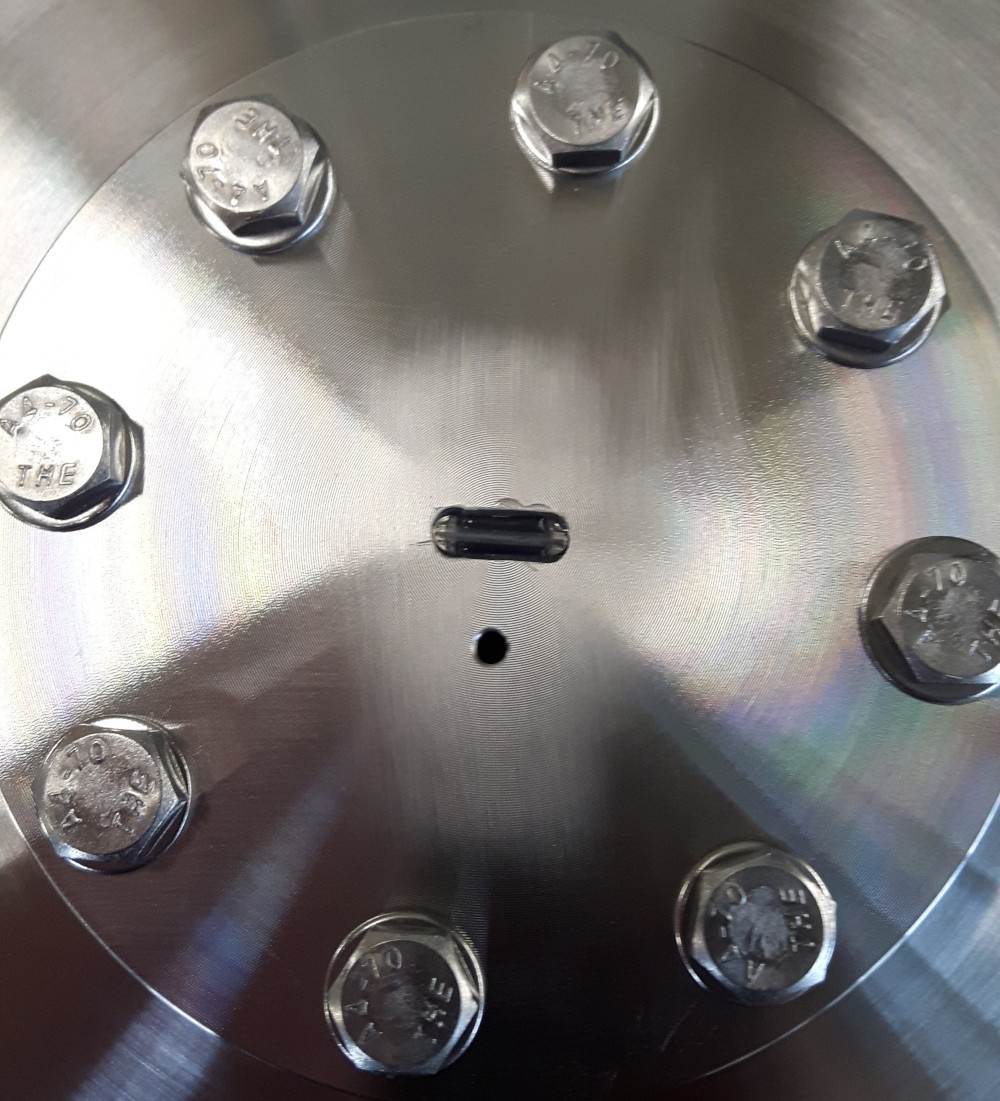}
        \caption{ }
        \label{fig:CF_flange}
    \end{subfigure}
    \caption{Channel sockets for \textbf{(a)} the IUSTI experimental facility and \textbf{(b)} the TRANSFLOW experimental facility.}
\end{figure}

For integrating the microchannels into the measurement setups, they are glued with the same epoxy into a socket with KF (ISO quick release) flanges on each side for the experiments at the IUSTI, see Fig. \ref{fig:KF_flange}, and into a CF (ConFlat, cooper-sealed) flange for the experiments at the KIT, see Fig. \ref{fig:CF_flange}.

\subsubsection{Functionalization}
\begin{figure}[ht]
\centering
\begin{subfigure}{.45\textwidth}
    \includegraphics[width=\linewidth]{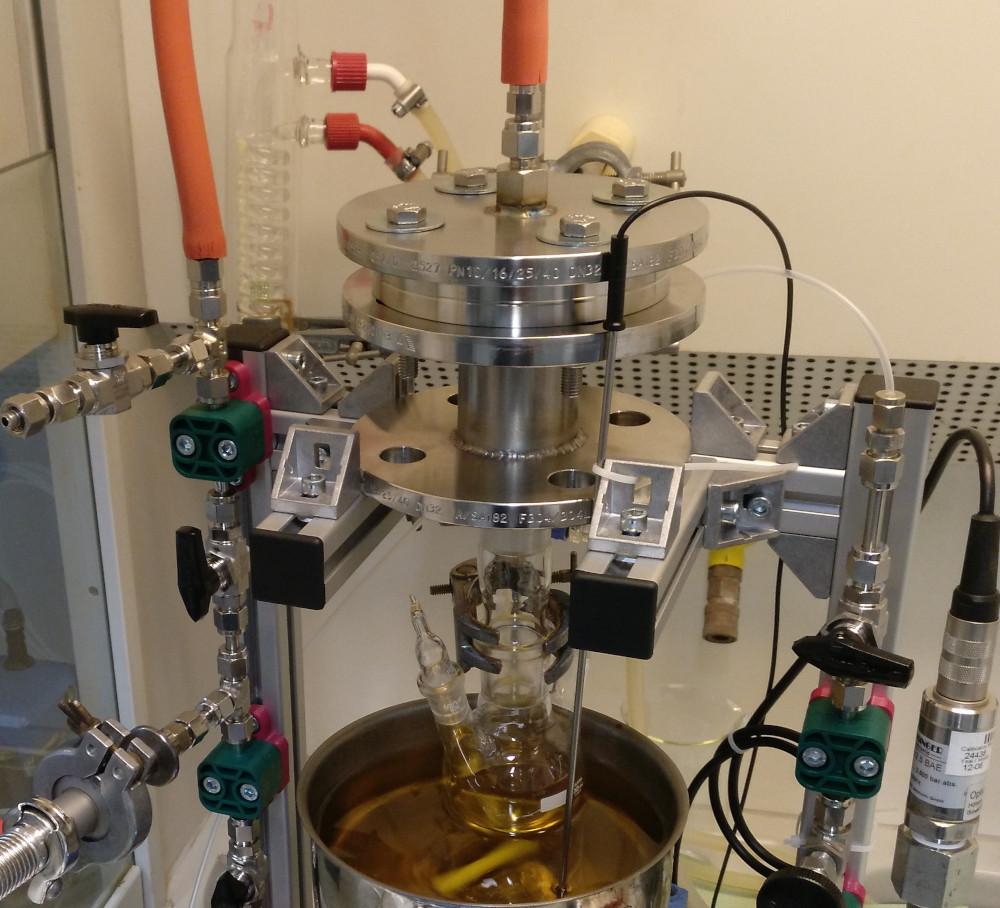}
    \caption{ }
    \label{fig:funct_setup}
\end{subfigure}
\hspace*{0.02\textwidth}
\begin{subfigure}{.45\textwidth}
    \includegraphics[width=\linewidth]{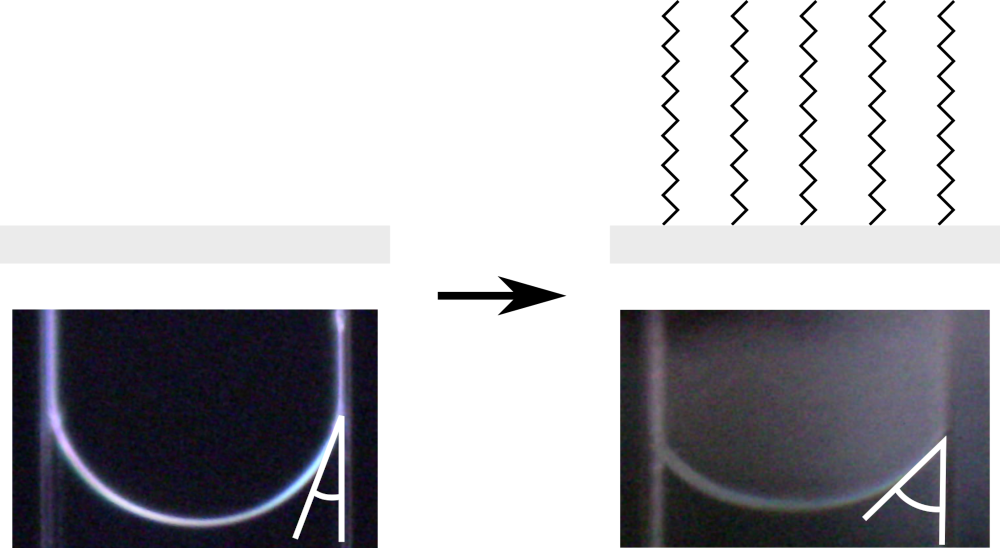}
    \caption{ }
    \label{fig:funct_contact_angle}
\end{subfigure}
\caption{\textbf{(a)}: The functionalization setup. \textbf{(b)}: Evaluation of functionalization by contact angle measurement (white) with water. Left: the plain channel. Right: the channel with HDTMS functionalization.}
\label{fig:functionalzation}
\end{figure}

The functionalization of the microchannels is performed by an in-house apparatus for chemical vapor deposition, see Fig. \ref{fig:funct_setup}. Only the small channels used at the IUSTI are functionalized, because in those the largest influence is expected. The channel socket attaches to the apparatus by KF flanges. The system is evacuated and flushed with nitrogen. After that, HDTMS is injected into the three-neck round-bottom flask with nitrogen excess pressure. While heating the flask in an oil bath to 150~$^\circ$C, the flask is being evacuated until the goal temperature is reached and condensate on the flask is visible. Then, vacuum is only pulled from the other side of the channel, creating a pressure difference along the channel and forcing the silane through it. This process runs for 62 hours. The system is then flooded with air, and the channels are extracted.

To evaluate the result of functionalization, the contact angle of water and air inside the channel is measured, with and without functionalization, see Fig. \ref{fig:funct_contact_angle}. For this, the microchannels are observed with an incident light microscope while water is inserted into the channel using a pipette. The results (16 $\pm$ 1.3° for plain, 36.3 $\pm$ 2.2° for functionalized) show a clear difference and therefore a successful functionalization. The analysis for the contact angle is performed by using ImageJ \cite{Schneider2012} with the contact angle plugin.

\subsection{Mass flow rate measurement facilities}
\label{subsec:facilities}
\begin{figure}[ht]
    \centering
    \includegraphics[width=6cm]{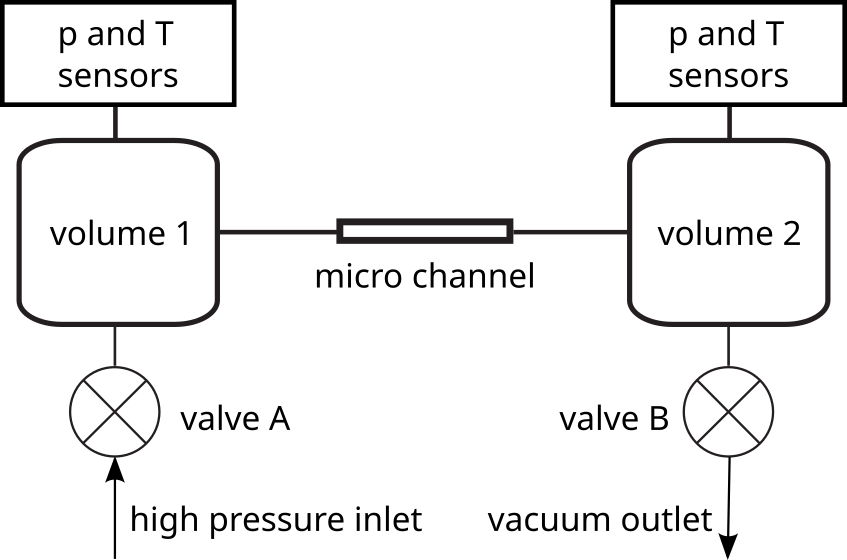}
    \caption{Schematic view of the experimental setups. During an experiment, valve B is closed for the IUSTI setup and open for the TRANSFLOW setup. Valve A is closed in both setups.}
    \label{fig:facilities}
\end{figure}

In both facilities, the constant volume method is used. This means that pressure changes in the upstream (and optionally downstream) reservoir are constantly measured, while the mass flow rate is deduced from the slope of the pressure evolution over time. The setups are schematically depicted in Fig. \ref{fig:facilities}.

\begin{figure}[ht]
\centering
\begin{subfigure}{.45\textwidth}
    \includegraphics[width=\linewidth]{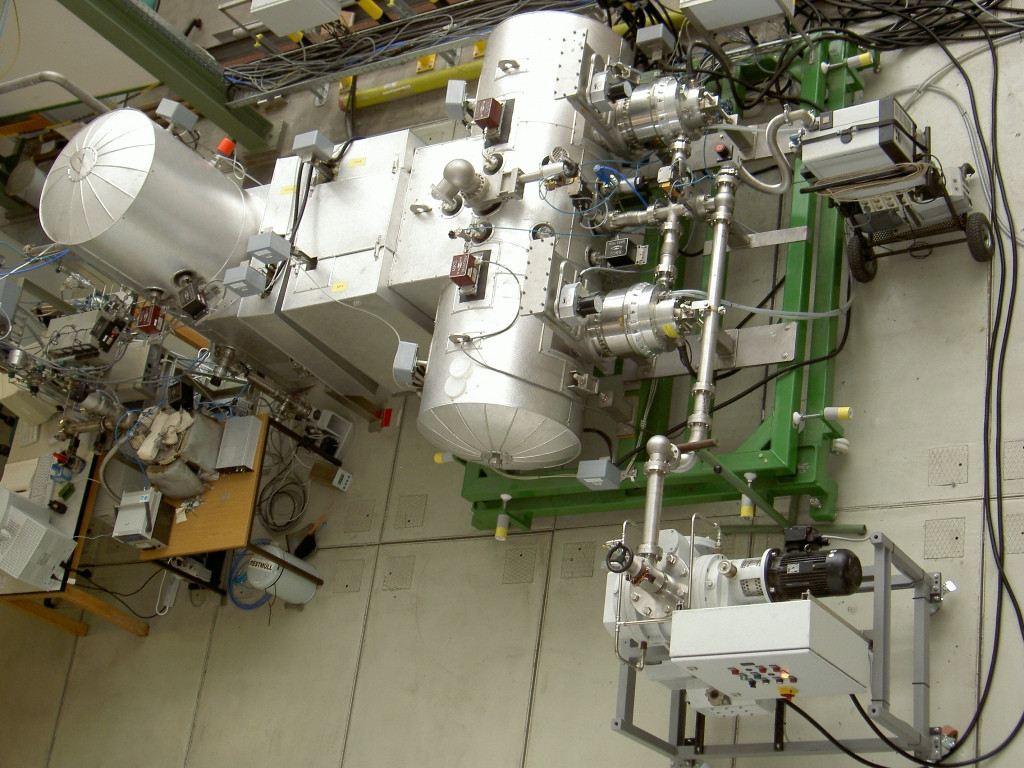}
    \caption{ }
    \label{fig:transflow}
\end{subfigure}
\hspace*{0.02\textwidth}
\begin{subfigure}{.45\textwidth}
    \includegraphics[width=\linewidth]{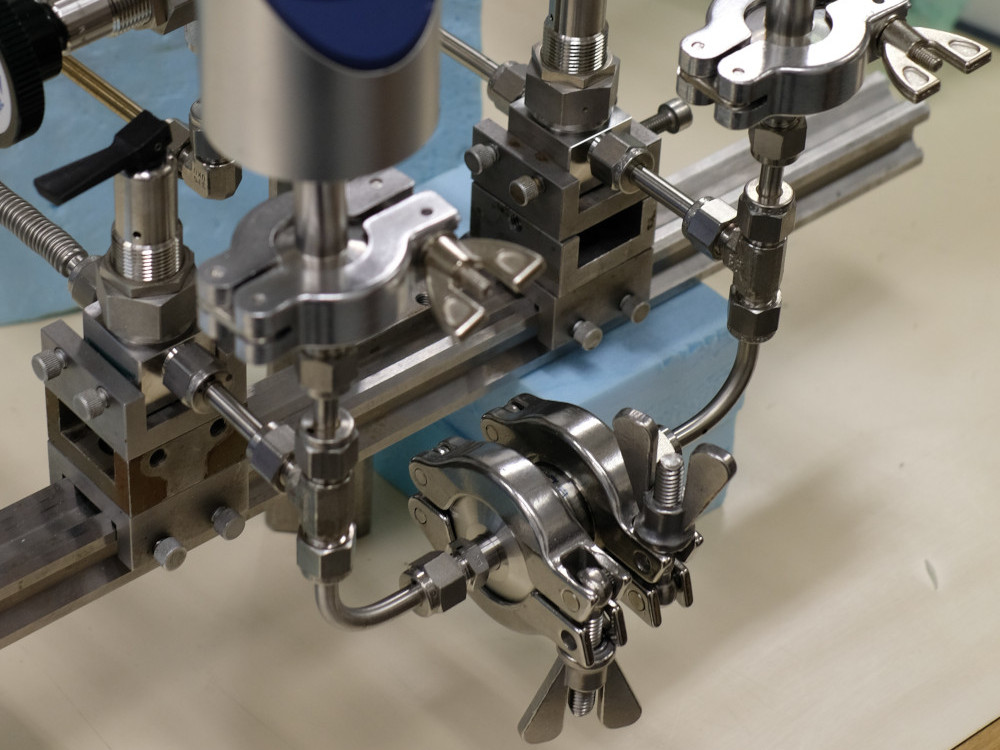}
    \caption{ }
    \label{fig:facility_iusti}
\end{subfigure}
\caption{\textbf{(a)}: Experimental facility TRANSFLOW. \textbf{(b)}: Experimental facility at the IUSTI laboratory.}
\end{figure}

In TRANSFLOW experimental facility, see Fig. \ref{fig:transflow}, the mass flow is measured by the constant volume method using only the upstream volume. The volume of the upstream reservoir including adapter flange is $V_1$ = 0.606 $\pm$ 0.012 $m^3$. Both reservoirs consist of 316LN stainless steel. All connected flanges are high vacuum CF flanges with copper sealing or Swagelok® connections. In the downstream reservoir, two turbomolecular pumps MAG W 2800 by Oerlikon-Leybold are attached via a VAT UHV gate valve. The volume of the downstream reservoir is $V_2$ = 1.2$m^3$ and is assumed quite large compared to the volume of the corresponding test channel. The whole facility can be heated with 6 heating circuits on the upstream, 7 on the downstream reservoir and one on the test channel. The temperatures can be adjusted within 1 degree.

To measure the pressure in both reservoirs, capacitance diaphragm gauges (CDG) by MKS, type Baratron® 690 HA, and one hot cathode gauge by Granville-Philips, type Stable Ion®, are installed on respectively. The CDGs provide a maximum measurement range of \SI{133 300}{\pascal}, \SI{1333}{\pascal} and \SI{133.3}{\pascal} at the dosing dome or \SI{13.33}{\pascal} at the pumping dome, respectively. Further details about the TRANSFLOW facility can be found in \cite{Varoutis2012}.

For the facility at the IUSTI laboratory, the constant volume method based on pressure measurements in both upstream and downstream volumes is used to determine the mass flow rate through a microchannel fixed between two volumes, see Fig. \ref{fig:facility_iusti}. A detailed description of the setup is given in \cite{Brancher2021}. The facility was slightly modified to work with the channel socket described in Section \ref{subsec:micro_channels}. To minimize the experimental time needed for a data point, the constant volumes 1 and 2, see Fig. \ref{fig:facilities}, are reduced and they consist merely of the piping. Those volumes are measured by pressure equilibration using a known third volume. The values of $V_1$=17.6 $\pm$ 0.4 cm$^3$ and $V_2$=17.2 $\pm$ 0.6 cm$^3$ are obtained for the upstream and downstream volumes, respectively.

During each experimental run the temperature is measured by K type thermocouples in combination with four-wire Pt100 for cold junction compensation. The temperature is stable within 3.1 K during an experimental run with a maximum length of around 17 hours and corresponds to room temperature.

With this setup it is possible to measure pressures in the range of 10 to \SI{100000}{\pascal} with the Capacitance Diaphragm Gauge (CDG). These sensors have an uncertainty of 0,2 \%. Two gases were used, helium and carbon dioxide, provided by Air Liquide (France) with a purity of 99.999 \%. The available pressure range corresponds to a Knudsen number range of 0.03 to 10 for helium and of 0.01 to 3 for carbon dioxide.

\subsection{Method for optimized mass flow rate measurement}\label{sec:MFR}

\subsubsection{General mass flow rate calculation}
When the pressure in the volumes is changing due to the mass flow through the channel and it does so slowly in relation to the time needed to reach a local equilibrium, we can assume that the system is in a quasi-stationary state. Therefore, under the quasi-stationary state assumption,  we can consider the change in the system as a succession of local equilibria. For a single gas and constant volume of a tank $V_i$ the change in mass of a gas inside is given by
\begin{equation}
 \diff m_i = m_i \left( \frac{\diff p_i}{p_i} - \frac{\diff T_i}{T}  \right).
\end{equation}
Over an infinitesimal time interval, we can calculate the mass flow rate as 
\begin{equation}
\frac{\diff m_i}{\diff t} =  \frac{V_i}{R T / M} \frac{\diff p_i}{\diff t} - \frac{p_i V_i}{R T^2 / M} \frac{\diff T}{\diff t}.
\end{equation}
The temperature in the tank is kept stable and nearly constant in time. Therefore, the change of temperature  in time compared to the change in pressure in time is very small, so we can consider the mass flow rate due to change of temperature in time negligible
\begin{equation}\label{eq:mfr_isotherm1}
\frac{V_i}{R T / M} \frac{\diff p_i}{\diff t} \gg \frac{p_i V_i}{R T^2 / M} \frac{\diff T}{\diff t},
\end{equation}
or equivalently, the relative change of pressure is much greater than the relative change in temperature 
\begin{equation}\label{eq:mfr_isotherm2}
\frac{\diff p_i}{p_i} \gg \frac{\diff T}{T}.
\end{equation}
Then, the mass flow rate, $\dot{m}$, is simply
\begin{equation}\label{eq:mfr_T}
\dot{m_i}(t)= \frac{\diff m_i}{\diff t} = \frac{V_i}{R T / M} \frac{\diff p_i}{\diff t}.
\end{equation}
If both valves A and B (see Fig. \ref{fig:facilities}) are closed, which is the case for the IUSTI facility but not for TRANSFLOW, the mass conservation between volume 1 and 2 results in
\begin{equation}\label{eq:mass_conservation}
\dot{m}_{1}(t)=-\dot{m}_{2}(t)
\end{equation}
using the convention of a negative mass flow for a reduction of mass inside a volume.

To calculate the mass flow rate the pressure variation in a tank over the time is measured. At the IUSTI laboratory, the pressure in both upstream and downstream volumes changes: the higher upstream pressure decreases over time while lower downstream pressure increases, until an equilibrium pressure, $p_{eq}$, is reached in both tanks. At the KIT laboratory, the downstream volume is continuously evacuated, therefore only the upstream pressure change is used for mass flow rate calculation.

These pressure variations in time can be fitted using either a linear function or more generally using an exponential function. To be able to cover a larger experimental range where the mass flow might not be constant, an exponential function is chosen here, as proposed in \cite{Cardenas2011}, \cite{Rojas-Cardenas2017}:
\begin{equation}
    \label{eq:exp_p}
    p_i(t) = p_{eq} + (p_i^*-p_{eq})e^{-t/\tau_i},
\end{equation}
where $p_{eq}$ is the equilibrium pressure for $t \rightarrow \infty$, $p_i^*$ is the initial pressure, and $\tau_i$ is the pressure relaxation time, which characterizes the speed of pressure rise or drop. The measured pressure variation in time is fitted using Eq. (\ref{eq:exp_p}) with $\tau$ and $p^*$ as the fitting parameters. $p_{eq}$ is approximated using the initial pressures and the tank volumina. The pressure derivation in time is calculated as
\begin{equation}\label{eq:exp_pdot}
\frac{\mathrm{d}p_i}{\mathrm{d}t} = -\frac{p_i^*-p_{eq}}{\tau_i} e^{-t/\tau_i}
\end{equation}
The mass flow rate is then calculated from
\begin{equation}\label{eq:mdot_from_p}
    \dot{m_i}(t) = \frac{V_i}{RT/M}\frac{p_i^*-p_{eq}}{\tau_i}e^{-t/\tau_i}.
\end{equation}
which simplifies for $t=0$ to
\begin{equation}\label{eq:mdot_from_p_t0}
    \dot{m_i} = \frac{V_i}{RT/M}\frac{p_i^*-p_{eq}}{\tau_i}.
\end{equation}
to get the mass flow rate at the beginning of the experiment. The Knudsen number associated to this mass flow is calculated using $p_m^* = 0.5 \left(p_1^*+p_2^*\right)$.

The measured mass flow rate is corrected with a previously measured leakage or outgassing mass flow rate by addition or subtraction when the leakage or outgassing is significant. The leakage mass flow rate is determined by setting the pressure equal in both volumes and by monitoring the pressure rise over time. The magnitude of the influence of the leakage correction varies strongly and is quantified in Appendix \ref{appendix:sec:leakages}.

Because the mass flow is calculated using the pressure drop in the volumes, the experiments need to run for a certain time to have a significant change in pressure compared to the sensor noise. However, too long measurement times introduce external influences like temperature fluctuation which can impact the temperature constancy assumption, required for implementing described mass flow rate extraction model, see Section \ref{sec:MFR}. Too short measurements in turn can lead to a high uncertainty due to the small amount of data available for a fit.

To find the ideal amount of data, the measurement uncertainty is minimized. The measurement uncertainty consists of two parts: one part describing the reliability of the chosen amount of measurement data using \textit{data fragments}, and the other part describing the physical uncertainties of pressure sensors, temperature fluctuations and channel and facility geometries.

\subsubsection{Variability of data fragments}
\label{subsubsec:data_fragments}
The \textit{variability of data fragments} methodology differs for the IUSTI and the TRANSFLOW facilities.

In case of the IUSTI facility, an important criterion for a successful run is the match of inlet and outlet mass flows, Eq. (\ref{eq:mass_conservation}), which are independently calculated using the respective pressure drops. This criterion is not fulfilled if the signal-to-noise ratio is too low, because that will introduce randomness to the mass flows and they likely will differ. Also, temperature fluctuations would create a mismatch because a temperature rise decreases the calculated mass flow rate in volume 1 and increases the calculated mass flow rate in volume 2.

\begin{figure}[ht] 
 \centering
    \includegraphics[width=8cm]{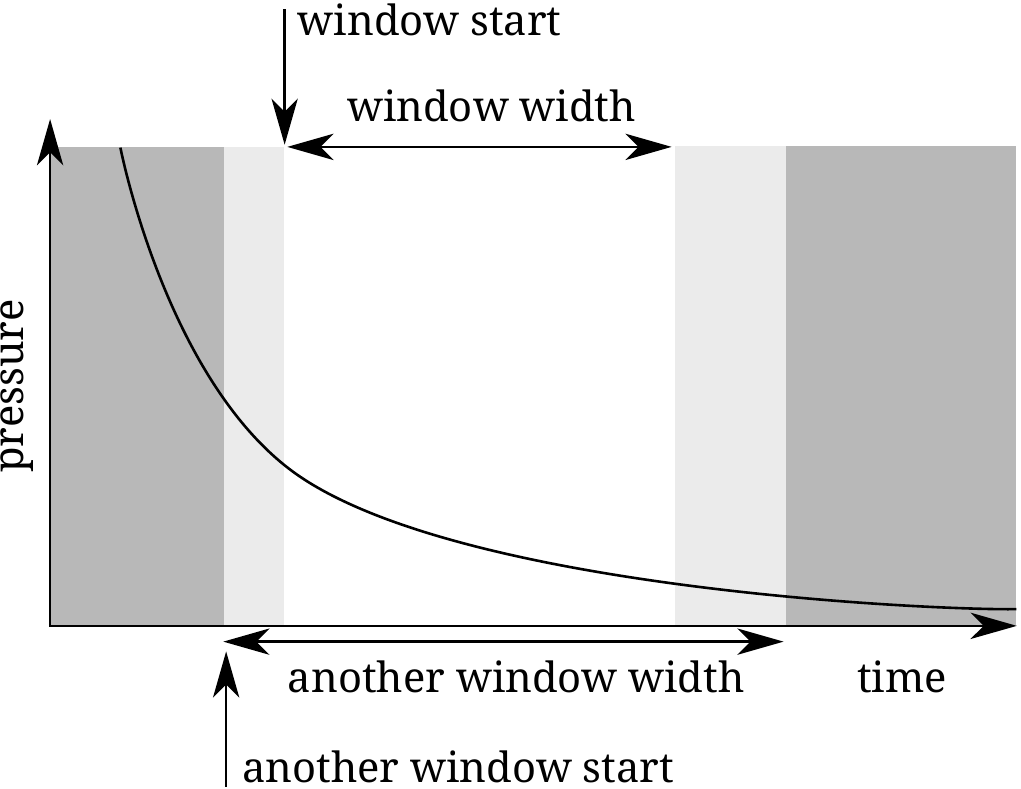}
    \caption{Depiction of two \textit{time windows}. A \textit{time window} is defined by a start and a width. One window consists of the data with white background. Another window consists of the data with white and light grey background: it has an earlier start and a larger width.}\label{fig:time-window}   
\end{figure}

\begin{figure}[ht]
\centering
\begin{subfigure}{.45\textwidth}
    \centering
    \includegraphics[width=\linewidth]{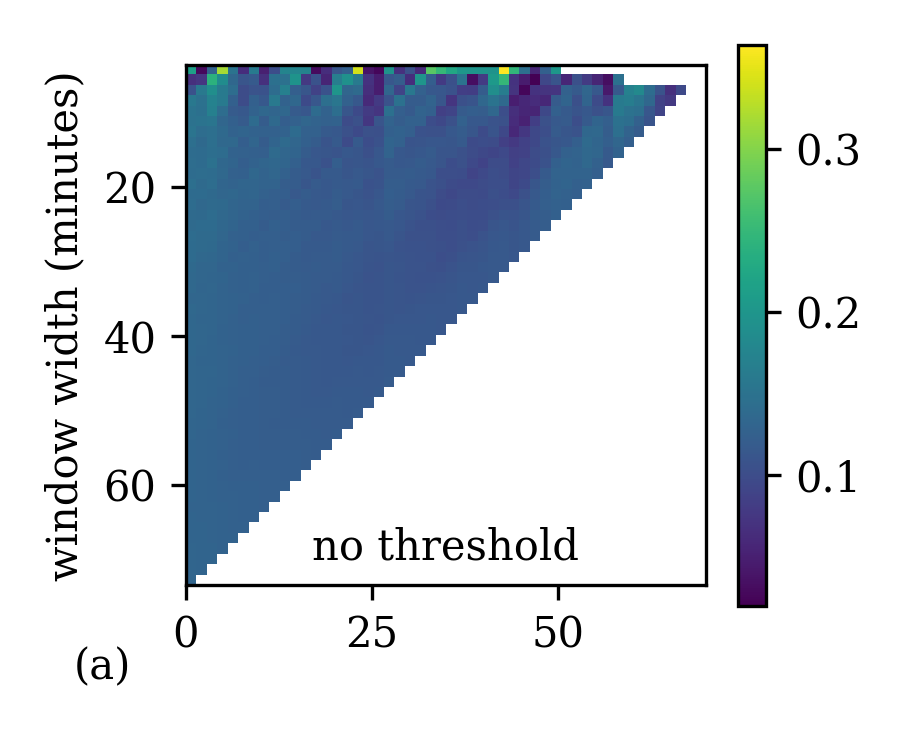}
    \vspace{-2em}
    \captionlistentry{}
    \label{fig:rel_diff_array_th1}
\end{subfigure}
\begin{subfigure}{.45\textwidth}
    \centering
    \includegraphics[width=\linewidth]{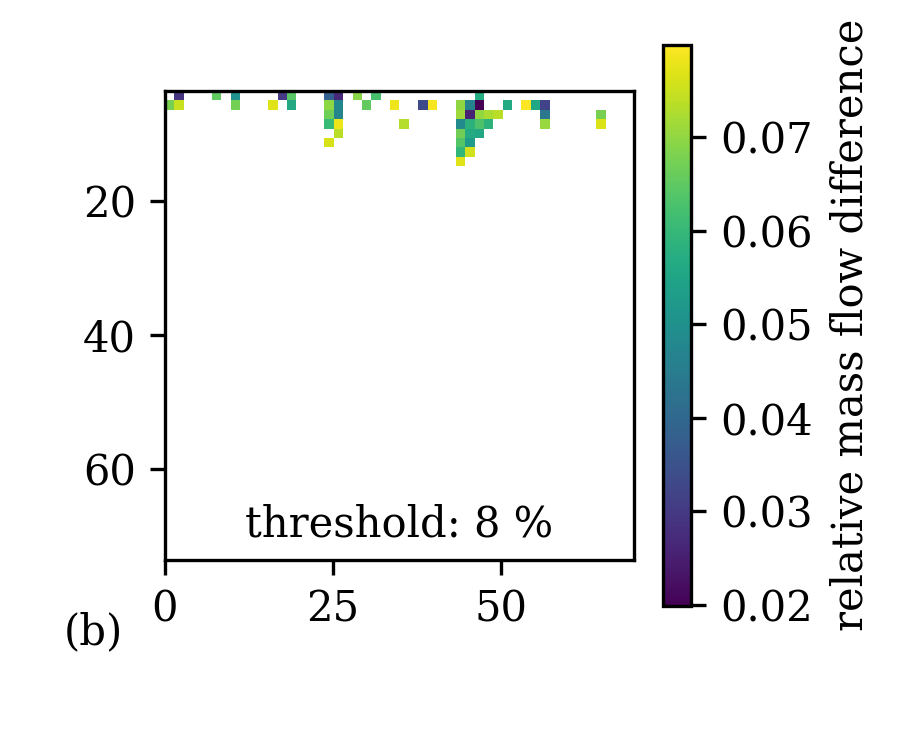}
    \vspace{-2em}
    \captionlistentry{}
    \label{fig:rel_diff_array_th2}
\end{subfigure}
\begin{subfigure}{.45\textwidth}
    \centering
    \includegraphics[width=\linewidth]{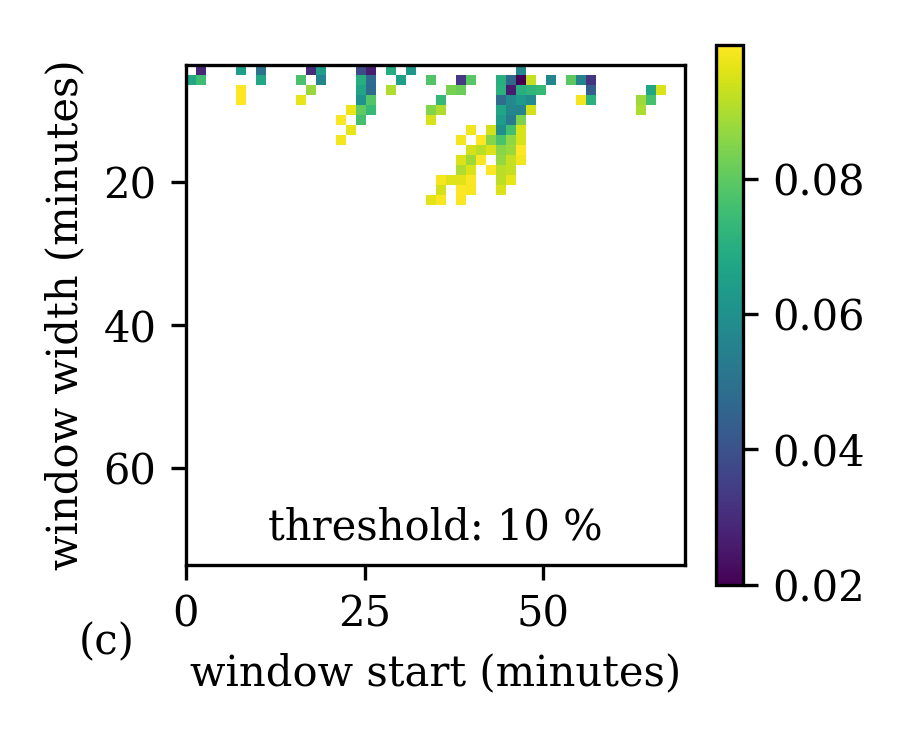}
    \captionlistentry{}
    \label{fig:rel_diff_array_th3}
\end{subfigure}
\begin{subfigure}{.45\textwidth}
    \centering
    \includegraphics[width=\linewidth]{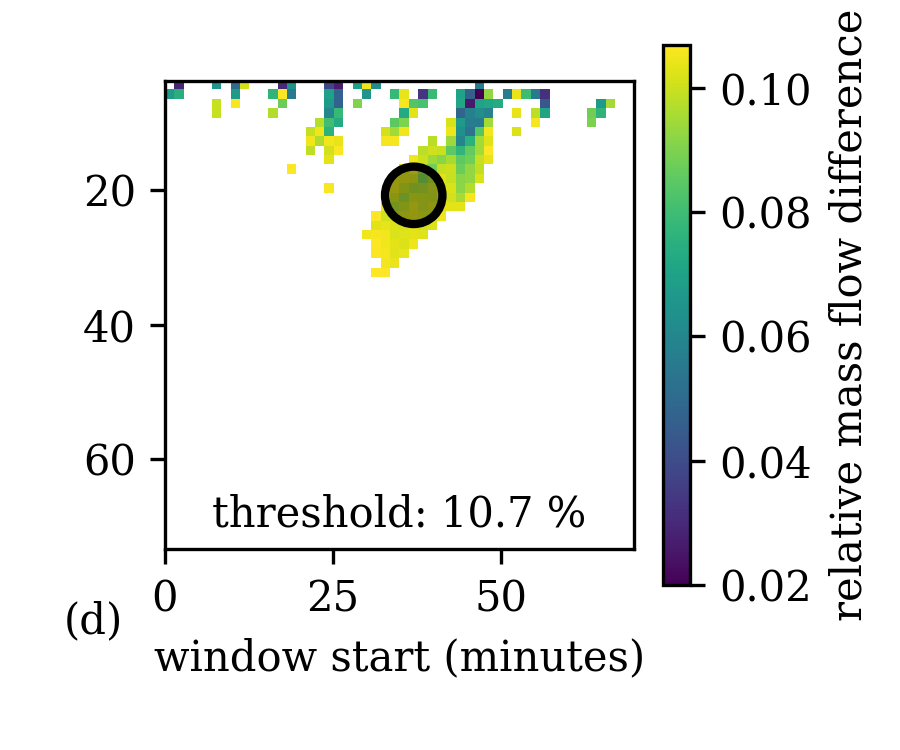}
    \captionlistentry{} 
    \label{fig:rel_diff_array_th4}
\end{subfigure}
\caption{Differences between mass flow rates calculated from upstream and downstream pressure changes. Each square corresponds to a fragment of the data defined by its corresponding window start and window width. The lower right part is empty because the window would end outside the available data. This results in (50x50)/2 = 1250 single evaluations. Some samples at the top are empty because no fit for that data was found. \textbf{(a)} The full array without any threshold. \textbf{(b)} A threshold of 8 \%; only very few fragments fulfill this threshold. No sufficiently large area can be found. \textbf{(c)} A threshold of 10 \%. More fragments fulfill this threshold, but not enough to find a sufficiently large area. \textbf{(d)} A threshold of 10.7 \% results in a structure where a sufficiently large area can be found (marked with a circle). Therefore, the corresponding relative uncertainty of mass flow rate due to the data fragment method is 5.35 \%.}
    \label{fig:massflow-array}
\end{figure}

Accordingly, the issues above can be addressed by looking at the difference of the mass flow rates calculated in volumes 1 and 2. To evaluate the correct experimental time, a single long experimental run is used. Only parts of this run are analyzed to simulate different lengths of experimental runs. These parts are taken at different positions and are of different length, resulting in \textit{time windows} as depicted in Fig. \ref{fig:time-window}. This results in a two-dimensional array of fictive experiments, each having their analyzed mass flow rates and their difference between volume 1 and volume 2. An example of a two-dimensional array, visualizing the relative difference between volume 1 and volume 2 mass flow rates, is shown in Fig. \ref{fig:rel_diff_array_th1}.

For the runs with small width, very low deviations are right next to very high deviations. This is the result of a low signal-to-noise ratio of little data. Simply choosing the data fragment having the smallest deviation would therefore be a matter of chance and not a reliable choice.

To tackle this problem, a second criterion is imposed: not only does a single fragment need to have a low difference between inlet and outlet, but an \textit{area} of fragments should stay below a certain deviation. This is done by applying a threshold to the array of mass flow differences and searching for an area large enough to host a circle with a radius of 20 \% of the width of the time window at the circle's center. This corresponds to a possible shift of the window start by 20 \% of the window width and a scaling of the window width by 20 \% while staying within the threshold. To avoid noise problems, a minimum of 3 time windows should be covered by the radius. To include as much data as possible for the fit, longer runs are prioritized if ambiguity arises. 

The threshold is increased until an area fulfilling the 20 \% criterion is found. That process is depicted in Fig. \ref{fig:rel_diff_array_th2} to \ref{fig:rel_diff_array_th4}. The Knudsen number and the dimensionless mass flow rate, defined by Eq. (\ref{eq:G}), are taken from the run in the center of the circle.

In case of TRANSFLOW, there is no difference between mass flow rates of volumes 1 and 2 because only volume 1 is considered, so the methodology is slightly different. Each entry of the array is investigated one by one. For each entry, the difference to all other fragments is calculated. This results in an array of mass flow differences - not between volumes 1 and 2, but between the current and all other data fragments. To this difference a threshold is applied like above and a check for an area fulfilling the 20 \% criterion is performed. If no area is found, the next entry of the array is investigated by re-calculating the difference to all other fragments, applying the threshold and checking for an area. If after cycling through all entries no area is found, the threshold is increased and all entries are investigated one by one again until an area is found.

This threshold is then used for an uncertainty which is defined via the population standard deviation \cite[p. 111]{Illowsky2018} as half of the relative difference between inlet and outlet mass flow rates or half of the relative difference between the current window and all other windows within the found area.

Generally, the method using two volumes should be preferred because the successful comparison of two separate mass flow balances is a strong indicator of correct measurements. Using only one volume leaves more room for systematic errors because influences like leakage or adsorption and desorption effects are not directly detectable.

\subsubsection{Monte Carlo simulations}
\label{subsubsec:montecarlo}
To quantify the influence of the pressure sensor uncertainties on the calculated pressure drop, a Monte Carlo simulation is performed. The sensor uncertainty is assumed to follow a normal distribution. For each sample, the error is calculated using the measured data and the sensor uncertainty, which is a constant factor for the IUSTI facility and a pressure-dependent factor according to calibration certificates for the TRANSFLOW facility.

This error is added to the measured data and a fit is performed. The pressure drop is calculated using this fit. This process is repeated with many samples until a convergence of the mean of the pressure drops resulting from the samples is reached. The standard deviation of the pressure drop between the samples is the uncertainty of the pressure drop due to the pressure sensor uncertainty.

\subsubsection{Error propagation}
\label{subsubsec:errorpropagation}
The influence of the uncertainties of pressure drop (calculated with the Monte Carlo method), temperature (calculated using the standard deviation of temperature during an experimental run), facility volumes (see Section \ref{subsec:facilities}) and channel geometry (calculated using the VSI data, see Section \ref{subsubsec:micro_channel_geometry_socket}) on the dimensional and dimensionless mass flow rates is determined using standard error propagation of independent variables. This is implemented using the Python package \texttt{uncertainties} \cite{Lebigot}.

\subsubsection{Minimization of uncertainties}
\label{subsubsec:optuncertainties}
The total uncertainty is the sum of the uncertainty coming from the data fragment analysis, see Section \ref{subsubsec:data_fragments}, and the uncertainty calculated by error propagation, see Section \ref{subsubsec:errorpropagation}. To minimize the total uncertainty, the best data fragment needs to be determined. This is done by sampling many thresholds for the array of mass flow differences and performing the complete calculation of the total uncertainty for each threshold. Thresholds with no valid 20 \% area are discarded. The threshold with the lowest total uncertainty is chosen, and the data fragment associated with this threshold is determined as the optimal data fragment. If this total uncertainty is still larger than 20 \%, that experimental run is discarded. This is by no means a polishing of data, but rather as a tool to identify experimental runs which have issues at the execution level, like excessive adsorption and desorption effects or leakage.

\subsection{Analytical and numerical solutions}
\label{subsec:analytical_numerical_methods}
For comparison with the experimental data, an analytical model and numerical  simulations are used. The analytical model, which was proposed in \cite{Kunze2022}, provides a mass flow rate expression and consists of a superposition of slip as well as diffusive flow. The latter is a combination of Fickian self-diffusion and free molecular flow. The mean free path, which is part of both the slip flow end the diffusive flow, is adjusted to take into account the channel geometry for high Knudsen numbers. For the molecular diameters of the gases, also influencing the mean free path, the transition molecular diameters as presented in \cite{Kunze2022} are chosen. The model is used without further modification for this work.

The relaxation type models of the Boltzmann equation collisional term, like BKG and S-model, are largely used for simulation of the gas flows through channels of different cross-sectional shapes \cite{Sharipov1998}. In this work, the linearized versions of both model kinetic equations, BGK and S-model, are solved numerically. This approach allows to simulate the gas flow through channels with a small ratio between the characteristic dimension of the cross-section and the channel length, $h/L\ll1$ (which is the case for the channels used in these experiments) and any pressure ratio between the tanks, see more details in \cite{Varoutis2009,Graur2014}. However, both models require additional information about the gas-surface interaction in terms of the accommodation coefficient \cite{Sharipov1998} which is set to 1 for this work. Furthermore, the fact that a polyatomic gas is treated as a monoatomic in the numerical simulations is fully justified from the fact that in an isothermal pressure driven flow, the internal degrees of freedom do not influence the flow rate through a duct \cite{Tantos2020}.

\section{Results and discussion}

\subsection{Measured mass flow rate}
The measured data is shown in a form of dimensionless mass flow rate, $G$, to exclude influence of geometric scale, gas species and temperature. This is done by dividing the measured mass flow rate, $\dot{m}$, by a mass flow rate often called \textit{Knudsen diffusion,}  \cite{Loeb1934} resulting in a non-dimensional mass flow rate of

\begin{equation}
    \label{eq:G}
    G = \dot{m} \frac{3 \Pi L}{8A^2 \Delta p} \sqrt{\frac{\pi RT}{2M}}
\end{equation}
where $\Pi$ is the channel perimeter, $L$ is the channel length, $A$ is the channel cross-sectional area, $\Delta p$ is the pressure difference between inlet and outlet tanks, $R$ is the universal gas constant, $T$ is the temperature and $M$ is the molar mass of the gas.

All experimental results are plotted as a function of the Knudsen number. The Knudsen number is defined using the microchannel height, the smallest dimension of the rectangular channel cross-section, as
\begin{equation}\label{eq:Kn}
Kn=\frac{\lambda}{h},
\end{equation}
where $\lambda$ is the molecular mean free path calculated by
\begin{equation}\label{eq:MFP}
\lambda=\frac{\mu(T)}{p_m} \sqrt{\frac{\pi RT}{2M}}
\end{equation}
where $\mu(T)$ is the temperature-dependent dynamic viscosity interpolated from data available in \cite{Lide1995} and $p_m=0.5(p_1+p_2)$ is the mean pressure between the inlet and outlet tanks, $p_1$ and $p_2$, respectively.

\begin{figure}[ht]
\centering
\begin{subfigure}{.45\textwidth}
    \includegraphics[width=\linewidth]{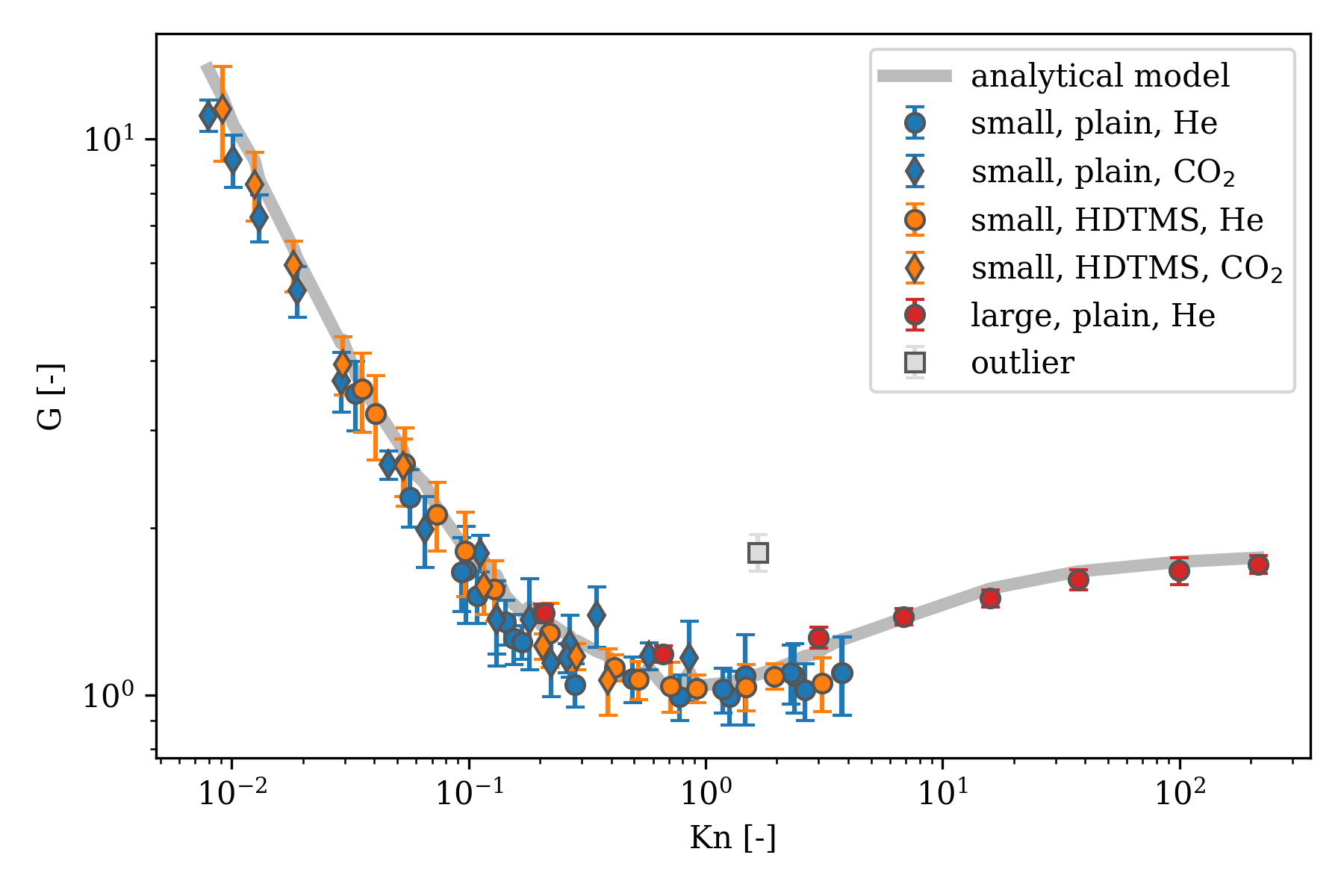}
        \caption{ }
    \label{fig:G-Kn-exp}
    \centering
\end{subfigure}
\begin{subfigure}{.45\textwidth}
    \includegraphics[width=\linewidth]{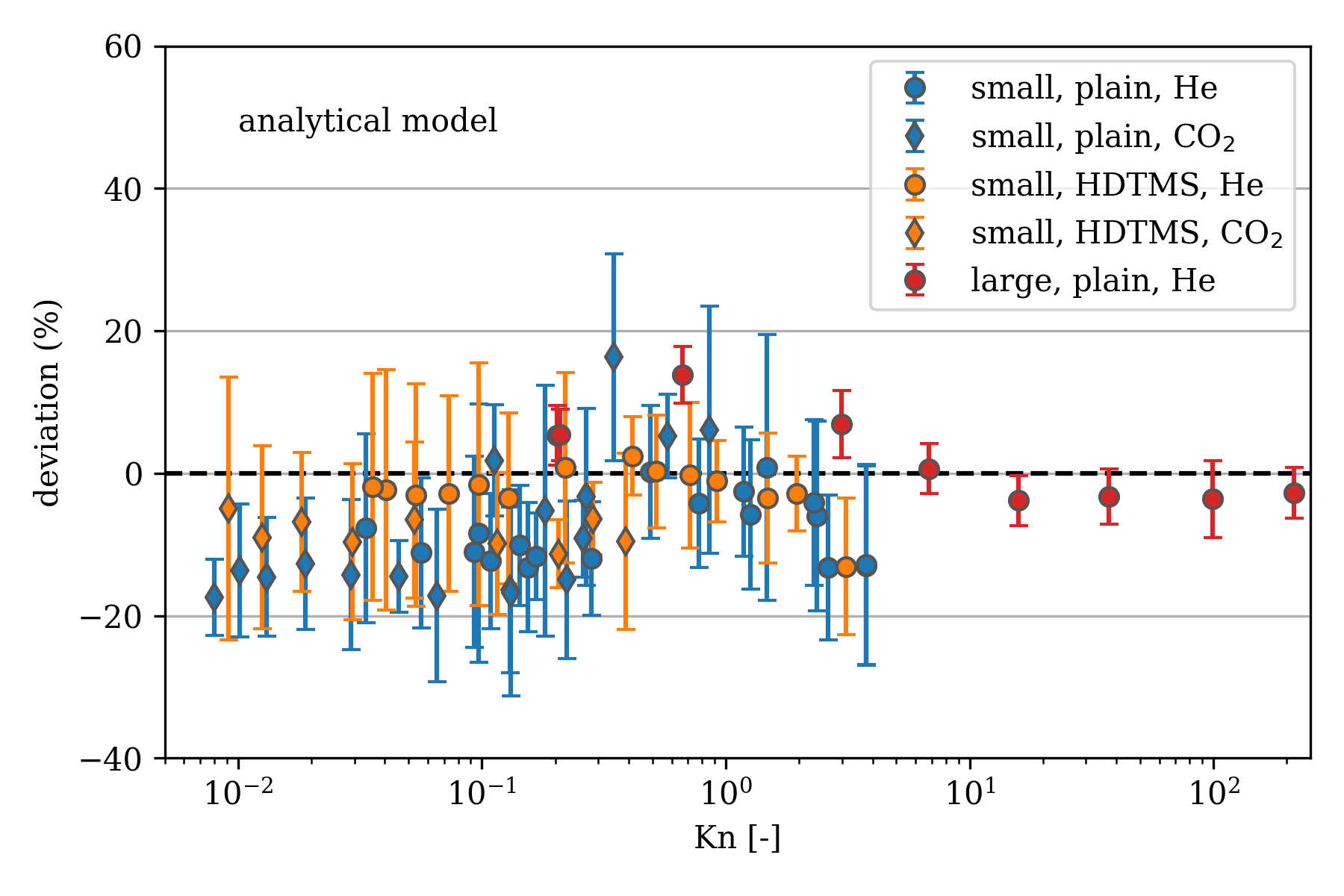}
    \caption{ }    
    \label{fig:G-Kn-exp-anal}
    \centering
\end{subfigure}
\vskip\baselineskip
\begin{subfigure}{.45\textwidth}
    \includegraphics[width=\linewidth]{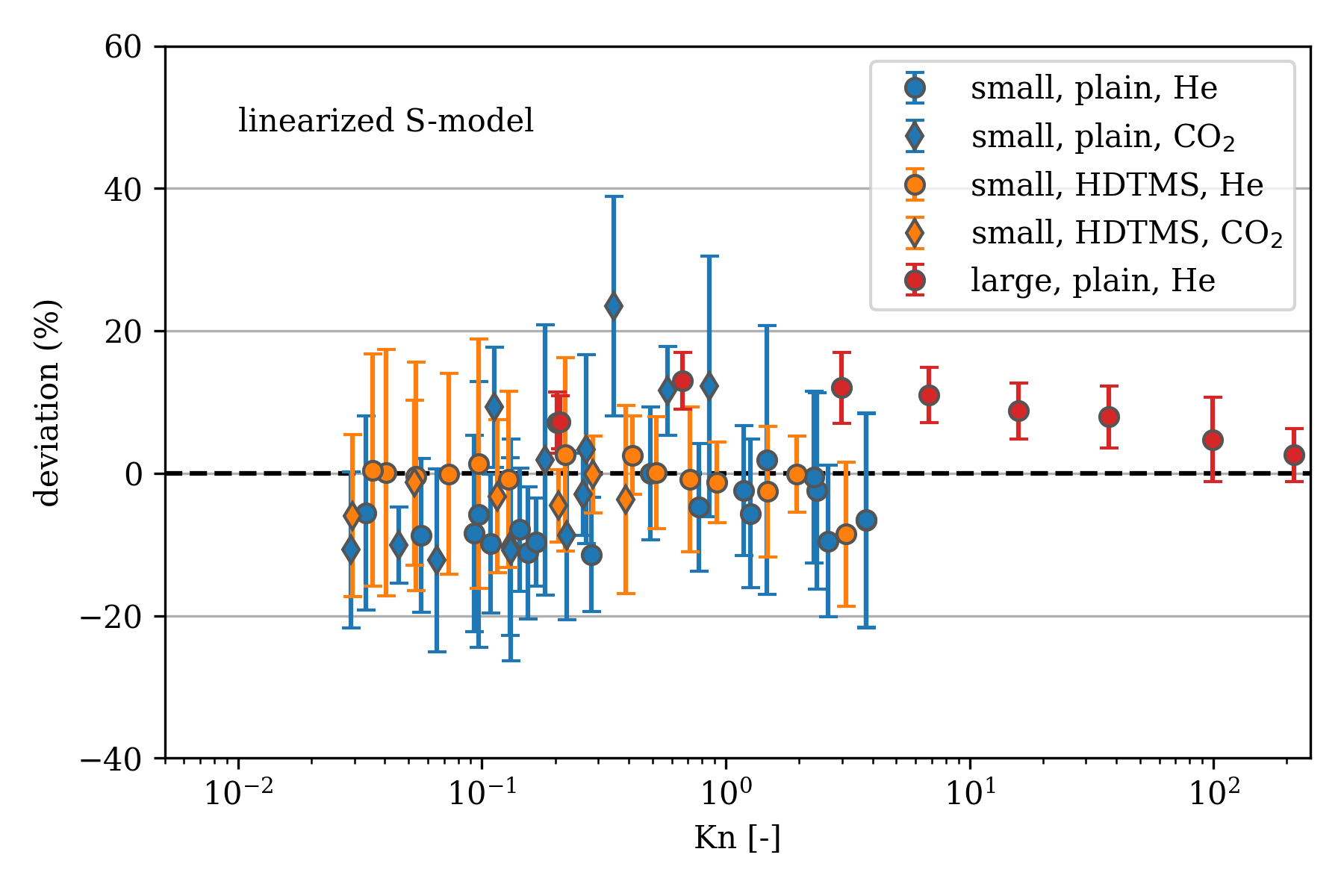}
    \caption{ }    
    \label{fig:G-Kn-exp-sim-graur}
    \centering
\end{subfigure}
\begin{subfigure}{.45\textwidth}
    \includegraphics[width=\linewidth]{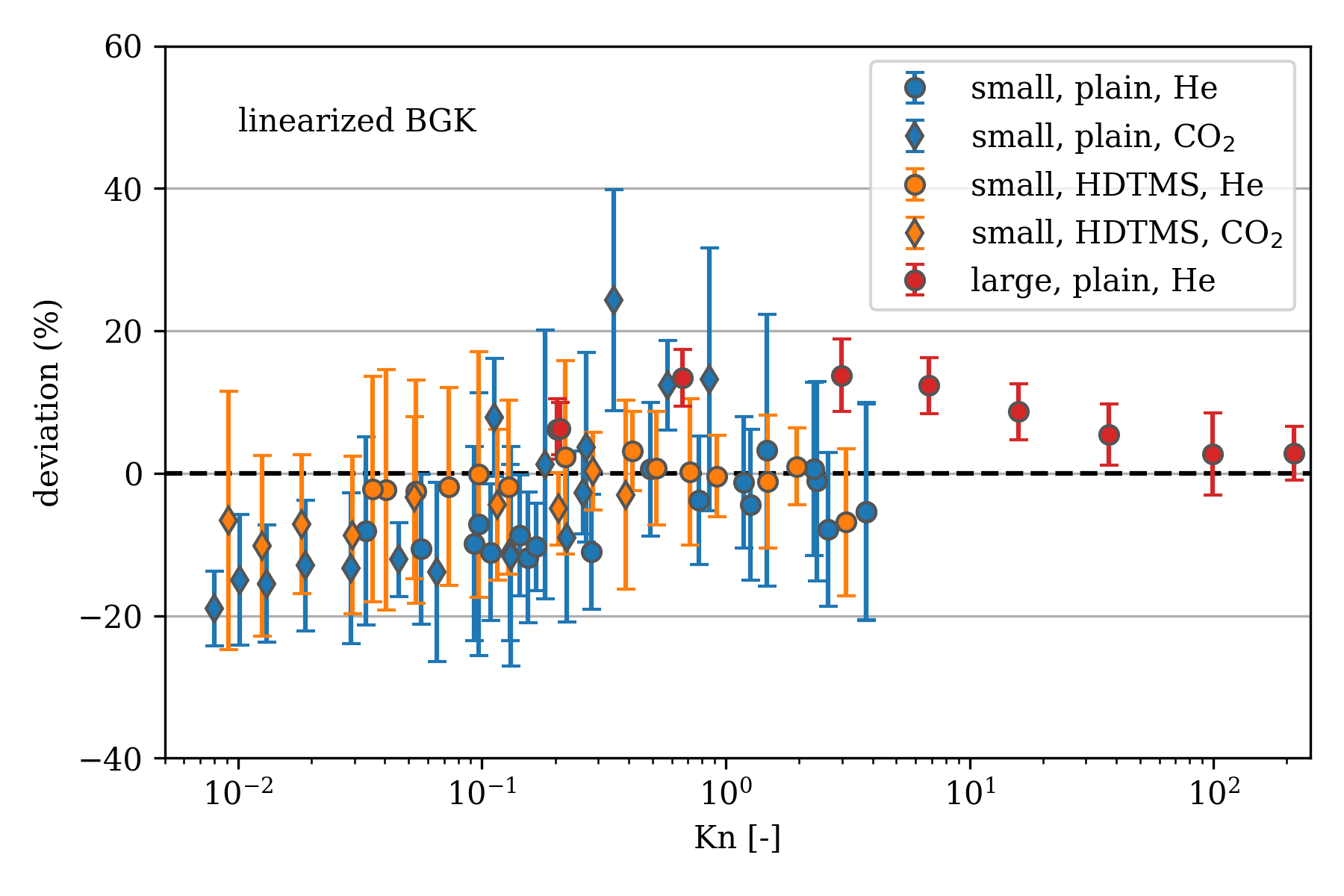}
    \caption{ }    
    \label{fig:G-Kn-exp-sim-varoutis}
    \centering
\end{subfigure}
\caption{\textbf{(a)} Dimensionless mass flow rate obtained experimentally (symbols) and calculated from the analytical model \cite{Kunze2022} (line). The outlier in grey color is not included in the Plots (b)-(d). \textbf{(b)}~Comparison of experiments to the analytical model. A negative value corresponds to an experimental value smaller than the analytical one. \textbf{(c)}~Comparison of experiments to the numerical solution of the linearized S-model equation~\cite{Graur2014}. \textbf{(d)}~Comparison of experiments to the numerical solution of the linearized BGK equation.} \label{fig:exp_data}
\end{figure}

The results of the experiments for both the plain and the HDTMS-functionalized channels with helium and carbon dioxide are shown in Fig. \ref{fig:G-Kn-exp} together with the analytical curve. The so-called Knudsen minimum \cite{Knudsen1909} at around Kn=1 is well visible. The dimensionless mass flow rate converges to a constant value for large Knudsen numbers as expected \cite{Beskok1999}. One experiment, although fulfilling the criterion of an uncertainty of $<$~20~\%, is considered an outlier and is therefore plotted in grey color and excluded from the Plots \ref{fig:G-Kn-exp-anal}-\ref{fig:G-Kn-exp-sim-varoutis} and further discussion.

It is clear from Fig. \ref{fig:G-Kn-exp} that HDTMS has no significant influence on the gas flow. This finding is different to previous measurements carried out in much smaller geometries \cite{Besser2020}. The absence of an impact of the functionalized surface may be explained by the fact that the influence of gas-surface interaction on the mass flow rate depends not only on the rarefaction level but also on the channel surface to volume ratio. This ratio for the channels used in experiments at the IUSTI laboratory is much smaller (0.4$\upmu$m$^{-1}$) than that of the previous experiments reported in \cite{Besser2020} (300$\upmu$m$^{-1}$, with a pore diameter of 20 nm). Therefore, the highest surface to volume ratio for the channels used in the experiments presented here is likely still too small to have any influence on the mass flow rate.

When comparing the experiments with the analytical model previously developed \cite{Kunze2022} (Fig. \ref{fig:G-Kn-exp-anal}) as well as with simulation data based on the linearized S-model kinetic equation from \cite{Graur2014} (Fig. \ref{fig:G-Kn-exp-sim-graur}) and that based on the linearized BGK kinetic equation (Fig. \ref{fig:G-Kn-exp-sim-varoutis}), all three results look very similar with little deviation between the analytical model and numerical solutions of both kinetic equations. All neighboring measurement points have overlapping measurement uncertainties, indicating an overall consistent result. All three models mostly lie within experimental error with deviation being smaller than 20~\% except for a single value.

To quantify the difference in mass flow rate between plain and functionalized channels, a t-test using unequal variances \cite{Welch1947} is performed. The populations are the deviations between experiment and analytical model for plain and functionalized channels, respectively. The null hypothesis is that the populations have the same means. The samples include all gases and channels. The test results in a p-value of 0.303, showing that there is no significant difference in mass flow rate between plain and functionalized channels.

\subsection{Numerical modeling using the linearized BGK-model equation}
\label{subsec:discussion_dvm_simulation}

The numerical solution of the BGK kinetic equation \cite{Varoutis2009} has been obtained
for the rectangular channel with high-to-width aspect ratio equal to 0.0328, which
correspond to the dimension of channels used in KIT experiments.

By comparing the corresponding numerical and experimental results, see Fig. \ref{fig:G-Kn-exp-sim-varoutis}, it can be observed that for monoatomic helium and for the case of the plain as well as the functionalized surface, a good agreement in the range of 0.03~$\leq$~Kn~$\leq$~110 is obtained, in which the relative error ranges between 1 and 14~\%. The magnitude of relative errors is evenly distributed across the Knudsen range. Moreover, from the current experimental data, no significant influence of the surface treatment is observed.

For the case of the polyatomic gas CO$_2$, it is observed that a good agreement in the range of 0.01~$\leq$~Kn~$<$~1 is achieved, with the relative error ranging between 2 and 25~\%, which is significantly higher than that for monoatomic helium. Large discrepancies are obtained in the case of Kn$\sim$1 and for Kn~$<$~0.1, which are mainly attributed to the high uncertainty of the experimental results. As in the case of helium, no significant influence of the surface treatment is observed. 

\subsection{Numerical modeling using S-model equation}
The authors of \cite{Graur2014} have performed numerical simulations of flow through channels of rectangular cross-sections with different aspect ratios using the linearized S-model kinetic equation. The channel cross-section aspect ratio used in experiments made at the IUSTI is equal to 0.0359. This value is very close to the aspect ratio 0.0367 for which the simulations have already been done by the authors of  \cite{Graur2014}. Therefore, the data for dimensionless mass flow rate from Table 1 of Ref. \cite{Graur2014} have been plotted in Figure \ref{fig:exp_data}(c) and very good agreement was obtained even for the polyatomic gas. A TMAC of 1 seems to describe the experimental data well for both the plain and the functionalized surfaces.

\subsection{Analytical model}
\label{subsec:discussion_analytical_model}
The analytical model developed previously by some of the authors \cite{Kunze2022} is used to calculate the mass flow rate under experimental conditions and is fully predictive. This means that the model is applied as described in \cite{Kunze2022} without any modification or adaption to the experimental data acquired in this work.

The model is based on extensions of existing expressions for the mass flow rate through channels of various cross-sections. Diffuse reflection is assumed when using the free molecular expression of the mass flow rate. This means that the tangential momentum accommodation coefficient TMAC, describing the ratio of diffuse to specular reflection, is equal to 1 for the free molecular regime. The TMAC is assumed to be equal to 0.9 when the slip flow expression is implemented. The authors of \cite{Kunze2022} fitted the model on available experimental data from literature using the molecular diameter as the only free parameter and obtained a so-called transition molecular diameter. The model was applied to the present data set implementing the previously calculated transition molecular diameters for helium and carbon dioxide, and very good agreement was found (Fig. \ref{fig:G-Kn-exp-anal}) between measured and calculated values of the mass flow rate.

While the model uses a TMAC of 0.9 for the slip flow in rectangular channels, in \cite{Kunze2022} it is stated that the actual TMAC should be 1 for all Knudsen numbers and the TMAC of 0.9 is attributed to a supposedly flawed slip expression for rectangular channels. The slip expression is derived from the solution of the Navier-Stokes equations using the Maxwell slip boundary condition and, for rectangular geometries, involves terms with convergent infinite series, while the expression for circular geometries is much simpler. There seems to be a complicating influence of the more complex rectangular geometry. A strong evidence that this is indeed the case is that if the TMAC was originally 0.9 for plain channels, it should become 1 when HDTMS is applied to the channel surfaces. HDTMS would sterically prevent any specular reflection, for which an even surface is needed \cite{Keerthi2018}. Because the mass flow does not change with functionalization, however, this means that the TMAC is 1 even for the plain channels. Both the calculations of the linearized S-model and BGK equations use a TMAC of 1, further strengthening this claim because of the good agreement with the experimental data.

\section{Conclusions}
The method developed to make optimal use of the inherently noisy mass flow rate data is suitable to yield results with reasonable uncertainties. While previous experiments in mesoporous structures show a clear influence of surface functionalization, the current experiments use much larger geometries. Here, the surface functionalization does not impact the mass flow rate through the channels considerably even though these experiments cover Knudsen numbers comparable to the experiments in mesoporous structures. This indicates that the Knudsen number is not the only parameter to consider but that there is a strong scaling dependence. The surface-to-volume ratio of the current channels is too small for the surface functionalization to significantly influence the mass flow rate.

The measured mass flow rate is compared to numerical solutions of the linearized BGK and S-model kinetic equations and with an analytical model showing good agreement in all cases. The fully diffusive gas-surface interaction seems to describe both types of channel surfaces well. The valid specification of TMAC = 0.9 for the slip expression of the analytical model indicates a potential for improvement in slip flow modeling of rectangular channels.

\section*{Acknowledgements}
This work was supported by German Research Foundation Grants TH~893/18-1 and TH~893/22-1.

\noindent Thanks to Christian Day for the support in this project.

\bibliographystyle{unsrt}
\bibliography{main}    

\appendix
\section{Leakages}
\label{appendix:sec:leakages}

\verb|D2_CO2_1000_100_leakage_07-12-21|
\begin{itemize}
    \item inlet sensor: 1000
    \item outlet sensor: 100
    \item $\dot{p}_{1}$: -5.915e-5 Pa/s
    \item $\dot{p}_{2}$: 3.561e-5 Pa/s
    \item used by:
    \begin{itemize}
        \item \verb|D2_CO2_1000_100_massflow_5k_1k_08-12-21|
        \item \verb|D2_CO2_1000_100_massflow_11k_2k_09-12-21|
        \item \verb|D2_CO2_1000_100_massflow_22k_4k_09-12-21|
        \item \verb|D2_CO2_1000_100_massflow2_11k_2k_09-12-21|
    \end{itemize}
\end{itemize}
\begin{figure}[h!]
    \centering
    \includegraphics[width=5cm]{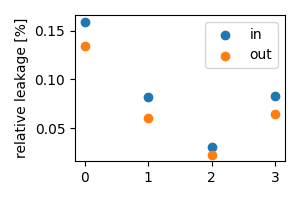}
    \cprotect\caption{Relative leakage of \verb|D2_CO2_1000_100_leakage_07-12-21|}  
\end{figure}

\noindent \verb|D2_He_1000_100_leakage_10k_14-01-22|
\begin{itemize}
    \item inlet sensor: 1000
    \item outlet sensor: 100
    \item $\dot{p}_{1}$: 0.0355 Pa/s
    \item $\dot{p}_{2}$: 0.00664 Pa/s
    \item used by:
    \begin{itemize}
        \item \verb|D2_He_1000_100_massflow_32k_12k_14-01-22|
        \item \verb|D2_He_1000_100_massflow_40k_11k_14-01-22|
        \item \verb|D2_He_1000_100_massflow_58k_11k_14-01-22|
    \end{itemize}
\end{itemize}
\begin{figure}[h!]
    \centering
    \includegraphics[width=5cm]{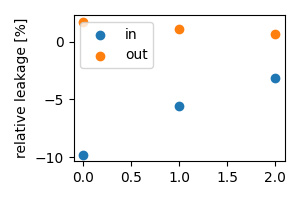}
    \cprotect\caption{Relative leakage of \verb|D2_He_1000_100_leakage_10k_14-01-22|}  
\end{figure}

\noindent \verb|D2_He_100_100_leakage_1700_18-01-22|
\begin{itemize}
    \item inlet sensor: 100
    \item outlet sensor: 100
    \item $\dot{p}_{1}$: 0.000945 Pa/s
    \item $\dot{p}_{2}$: 0.000973 Pa/s
    \item used by:
    \begin{itemize}
        \item \verb|D2_He_100_100_massflow_5200_1100_18-01-22|
        \item \verb|D2_He_100_100_massflow_6k_3k_18-01-22|
        \item \verb|D2_He_100_100_massflow_10k_5k_18-01-22|
    \end{itemize}
\end{itemize}
\begin{figure}[h!]
    \centering
    \includegraphics[width=5cm]{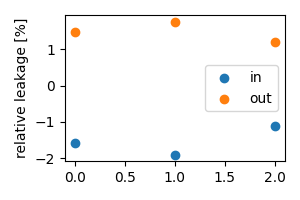}
    \cprotect\caption{Relative leakage of \verb|D2_He_100_100_leakage_1700_18-01-22|}  
\end{figure}

\noindent \verb|D2_He_1000_1000_leakage_42k_08-02-22|
\begin{itemize}
    \item inlet sensor: 1000
    \item outlet sensor: 1000
    \item $\dot{p}_{1}$: 0.0129 Pa/s
    \item $\dot{p}_{2}$: 0.00485 Pa/s
    \item used by:
    \begin{itemize}
        \item \verb|D2_He_1000_1000_massflow_133k_90k_08-02-22|
        \item \verb|D2_He_1000_1000_massflow_83k_49k_08-02-22|
        \item \verb|D2_He_1000_1000_massflow_56k_25k_08-02-22|
    \end{itemize}
\end{itemize}
\begin{figure}[h!]
    \centering
    \includegraphics[width=5cm]{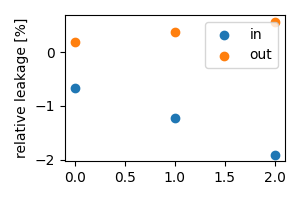}
    \cprotect\caption{Relative leakage of \verb|D2_He_1000_1000_leakage_42k_08-02-22|}  
\end{figure}

\noindent \verb|C2_He_100_100_leakage_700_18-03-22|

\begin{itemize}
    \item inlet sensor: 100
    \item outlet sensor: 100
    \item $\dot{p}_{1}$: 0.000861 Pa/s
    \item $\dot{p}_{2}$: 0.00115 Pa/s
    \item used by:
    \begin{itemize}
        \item \verb|C2_He_100_100_massflow_12k_3k_18-03-22|
        \item \verb|C2_He_100_100_massflow_13k_5k_18-03-22|
        \item \verb|C2_He_100_100_massflow_6500_1600_18-03-22|
        \item \verb|C2_He_100_100_massflow_7900_2600_18-03-22|
        \item \verb|C2_He_100_100_massflow_1950_400_18-03-22|
        \item \verb|C2_He_100_100_massflow_3400_430_18-03-22|
        \item \verb|C2_He_100_100_massflow_4300_700_18-03-22|
    \end{itemize}   
\end{itemize}
\begin{figure}[h!]
    \centering
    \includegraphics[width=5cm]{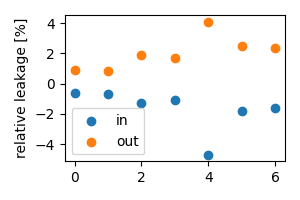}
    \cprotect\caption{Relative leakage of \verb|C2_He_100_100_leakage_700_18-03-22|}  
\end{figure}

\noindent \verb|C2_CO2_100_100_leakage_14-03-22|

\begin{itemize}
    \item inlet sensor: 100
    \item outlet sensor: 100
    \item $\dot{p}_{1}$: 0.00192 Pa/s
    \item $\dot{p}_{2}$: 0.00217 Pa/s
    \item used by:
    \begin{itemize}
        \item \verb|C2_CO2_100_100_massflow_5k_1k_14-03-22|
        \item \verb|C2_CO2_100_100_massflow_6k_2k_14-03-22|
        \item \verb|C2_CO2_100_100_massflow_12k_3k_14-03-22|
        \item \verb|C2_CO2_100_100_massflow_12k_3k_14-03-22|
    \end{itemize}
\end{itemize}
\begin{figure}[h!]
    \centering
    \includegraphics[width=5cm]{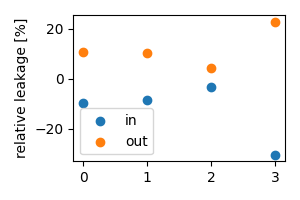}
    \cprotect\caption{Relative leakage of \verb|C2_CO2_100_100_leakage_14-03-22|}  
\end{figure}

\end{document}